\documentclass[superscriptaddress,twocolumn,showpacs,amsmath,amssymb]{revtex4}
\usepackage{graphicx}

\def\be{\begin{equation}}
\def\ee{\end{equation}}
\def\bea{\begin{eqnarray}}
\def\eea{\end{eqnarray}}
\def\br{{\bf r}}
\renewcommand{\vec}[1]{\mbox{\boldmath $#1$}}

\begin{document}

\title{
Diproton correlation 
in a proton-rich Borromean nucleus 
$^{17}$Ne }

\author{T. Oishi}
\author{K. Hagino}
\affiliation{
Department of Physics,  Tohoku University,  Sendai,  980-8578,  Japan}

\author{H. Sagawa}
\affiliation{
Center for Mathematical Sciences,  University of Aizu,  Aizu-
Wakamatsu,  Fukushima 965-8560,  Japan}


\begin{abstract}
We carry out three-body-model calculations for a proton-rich 
Borromean nucleus $^{17}$Ne by 
assuming a $^{15}$O + $p$ + $p$ structure. 
To this end, 
we use a density-dependent contact interaction between the 
valence protons, explicitly treating also 
the Coulomb interaction. 
We find that the two-particle density distribution for 
$^{17}$Ne is similar to that for 
$^{16}$C, which has two valence neutrons outside the $N$ = 8 core. 
That is, the two protons 
take a spatially compact configuration, 
while the Coulomb repulsion plays a minor role.
This indicates that there is a strong diproton correlation 
in the ground state of the 
$^{17}$Ne nucleus. 
We also show that 
the Coulomb interaction 
reduces 
the expectation value of the proton-proton interaction 
by about 14\%. 
\end{abstract}


\maketitle

\section{introduction}

Physics of unstable nuclei have attracted much attention both 
experimentally and theoretically \cite{T95,J04,HJ87}. 
These nuclei are located far from the $\beta$-stability line, 
having a large asymmetry in the proton-to-neutron ratio. 
In the neutron-rich side, 
a few exotic features have been observed. 
These include 
i) a large extension of density distribution, referred to as halo or skin 
structure\cite{T85}, ii) a narrow momentum 
distribution\cite{K88}, and iii) a large concentration of the dipole strength 
distribution at low energies\cite{F04,A99,N06}. 
For $^{11}$Li and $^6$He nuclei, the so called Borromean 
structure has also been discussed extensively \cite{BE91,Zhukov93}. 
The Borromean is defined as a 
three-body bound system in which any two-body subsystem does not 
bound. 
The pairing interaction between the valence neutrons plays an 
essential role in stabilizing these nuclei \cite{BE91}. 

One of the topics of current interest in physics of unstable nuclei 
is dineutron correlations in neutron-rich nuclei. 
Although two neutrons do not form a bound state in vacuum, 
theoretical calculations have shown that 
they take a spatially compact configuration in finite nuclei 
\cite{HJ87,BE91,Zhukov93,HS05,HSCS07,MMS05,PSS07,M06}. 
This feature is enhanced significantly 
when the binding is weak as in 
$^{11}$Li and $^6$He nuclei. 
The recent experimental observation of
the strong low-lying dipole strength distribution in the $^{11}$Li
nucleus \cite{N06} has provided an
experimental signature of 
the existence of strong dineutron correlation in this nucleus. 

In this paper, we discuss whether there exists a similar 
correlation in proton-rich nuclei. 
In contrast to the neutron-rich nuclei, proton-rich nuclei have been 
investigated less extensively, and the existence of diproton correlation has 
not yet been fully clarified. 
For this purpose, we choose $^{17}$Ne nucleus. 
This nucleus is known to be 
a candidate of two-proton halo nucleus, as has been 
suggested by measured large interaction cross sections \cite{Oz94} 
as well as a recent measurement of charge radius \cite{Gei08}. 
This nucleus is also known to have a Borromean nature, as its subsystem 
$^{16}$F is proton unbound. 
The two-proton emission decay 
has been observed from the first excited state 
of $^{17}$Ne nucleus\cite{C97,C02,Z04}, which suggests that the structure of 
$^{17}$Ne can be understood as a three-body system of 
$^{15}$O+$p+p$. Based on this idea, several three-body model calculations 
have been performed for $^{17}$Ne \cite{ZhTh95,TDB96,GMZ03,GFJ04,GLSZ06,GPZ05}. 
In this paper, we shall carry out similar three-body model 
calculations as 
in Refs. \cite{HS05}, in which the dineutron correlation in $^{11}$Li 
and $^6$He has been investigated. 
This is based on a three-body model with a density-dependent pairing 
interaction between the valence nucleons \cite{BE91,EBH97}. 
For the $^{17}$Ne nucleus, we also include the Coulomb interaction 
explicitly and discuss the role of the Coulomb repulsion in diproton 
correlation. 

The paper is organized as follows. 
In Sec. II, we set up the three-body model with a density-dependent 
contact interaction. 
In Sec. III, we apply the model to 
the ground 
state of $^{17}$Ne. We study the density distribution of the valence 
protons and discuss the diproton correlation. 
We also compare our results with those for the $^{16}$C nucleus, 
whose neutron structure $(N=10)$ 
may have some similarity to the proton structure $(Z=10)$ in 
$^{17}$Ne. 
We then summarize the paper in Sec. IV.

\section{Three-body model}

We describe the ground state of $^{17}$Ne nucleus 
by assuming an inert core nucleus $^{15}$O and two valence protons. 
We consider the
following three-body Hamiltonian for this system:
\bea
 H({\bf r}_1,{\bf r}_2)
 &=& h^{(1)}+h^{(2)}+\frac{{\bf p}_1 \cdot {\bf p}_2}{A_{\rm c}m}
    +V_{\rm pp}({\bf r}_1,{\bf r}_2), \label{eq:3bh} \\
 h^{(i)}
 &=& \frac{{\bf p}_i^2}{2\mu}+V_{\rm pc}(r_i), \label{eq:sph}
\eea
where $m$ and $A_{\rm c}$ are the nucleon mass and the mass number of
the inert core nucleus, respectively.
$h$ is the single-particle (s.p.) Hamiltonian for a valence
proton, in which $V_{\rm pc}$ is the potential 
between the proton and the core nucleus. 
$V_{\rm pp}$ is the interaction between the valence protons. 
The diagonal component of the recoil kinetic energy of the core
nucleus is included in the s.p. Hamiltonian $h$ through 
the reduced mass $\mu=A_{\rm c}m/(A_{\rm c}+1)$, 
whereas the off-diagonal part is
taken into account in the last term in Eq. (\ref{eq:3bh}) \cite{EBH97,HS05}. 

For the single-particle potential 
$V_{\rm pc}$, 
we use a Woods-Saxon potential together with the Coulomb interaction 
between a valence proton and the core nucleus. 
That is, 
\begin{equation}
 V_{\rm pc}(r)
 = \left[ V_0 + V_{ls}r_{0}^2(\vec{\ell} \cdot \vec{s})
     \frac{1}{r} \frac{d}{dr} \right] f(r) + V_{\rm C}(r), 
\label{eq:WSC}
\end{equation}
with
\begin{equation}
 f(r) = \frac{1}{1+\exp \left[ (r-R_{\rm core})/a_{\rm core})
          \right] },
\ee
where $R_{\rm core}=r_0A_{\rm c}^{1/3}$ is the radius 
of the core nucleus and $V_{\rm C}$ is the Coulomb 
potential for a uniform charge distribution. 
For the parameters, we use $r_0$=1.22 fm, $a_{\rm core}$=0.65 fm, 
$V_0=-53.73$ MeV, and $V_{ls}=15.01$ MeV. 
This parameter set yields an s$_{1/2}$ and a d$_{5/2}$ resonances 
at 0.675 MeV and 1.129 MeV, respectively. 
These are spin-averaged energies of the observed 
0$^-$ resonance at 0.535 MeV and 1$^-$ resonance at 0.722 MeV, and 
2$^-$ at 0.951 MeV and 3$^-$ at 1.257 MeV in $^{16}$F, 
respectively\cite{C02} (see Fig. 1). 

\begin{figure}[t]
\label{fig:fig1}
\begin{center}
\includegraphics[width=7.0cm, height=5.0cm]{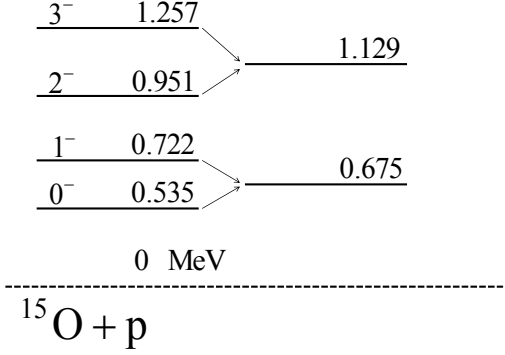}
\caption{A low-lying level scheme of 
$^{16}$F nucleus. 
Spin averaged levels for two sets of resonance states are also shown.}
\end{center}
\end{figure}

We solve the three-body Hamiltonian, Eq. (\ref{eq:3bh}), by expanding the 
ground state wave function 
with the single-particle wave functions, 
$\phi_{nljm}({\bf r})$\cite{BE91,HS05,EBH97}, 
\bea
 \Psi_{\rm g.s.}({\bf r}_1,{\bf r}_2)
 &=& \sum_{n \leq n'} \sum_{l,j} \alpha_{nn'lj} \tilde{\psi}_{nn'lj}
     ({\bf r}_1,{\bf r}_2), \\
 \tilde{\psi}_{nn'lj}({\bf r}_{1},{\bf r}_{2})
 &=& \nonumber \frac{1}{\sqrt{2(1+\delta_{nn'})}} \sum_{m} \left< j,m;j,-m \mid 0,0 \right> \\
 & & \nonumber \times \left[ \right.
     \phi_{nljm}(\br_1)\phi_{n'lj-m}(\br_2) \\
 & &+\phi_{n'ljm}(\br_1)\phi_{nlj-m}(\br_2) \left. \right], 
\label{eq:0+sp}
\eea
where we have assumed that the valence protons form a 
$0^+$ configuration. 
The continuum s.p. states are discretized in a large box with the size of 
$R_{\rm box}$=30 fm.
We include the s.p. angular momentum $l$ up to 12.
In the expansion, we explicitly exclude the
$1s_{1/2}$, $1p_{3/2}$, and $1p_{1/2}$ states, which are occupied
by the protons in the core nucleus.

In this work, we employ a density-dependent
contact interaction for the nuclear part of the proton-proton interaction 
\cite{BE91,EBH97,VMP96,HS05}, 
while the Coulomb potential is also explicitly included. 
That is, 
\be
V_{\rm pp}({\bf r}_1,{\bf r}_2) = 
V^{(N)}_{\rm pp}({\bf r}_1,{\bf r}_2) + 
V^{(C)}_{\rm pp}({\bf r}_1,{\bf r}_2), 
\ee
with 
\begin{equation}
V^{(N)}_{\rm pp}({\bf r}_1,{\bf r}_2) = 
\delta(\br_1-\br_2) \left[ v_0+\frac{v_{\rho}}{1+\exp
     \left[(r_1-R_{\rho})/a_{\rho} \right]} \right], 
\label{eq:Vpp}
\ee
and
\be
V^{(C)}_{\rm pp}({\bf r}_1,{\bf r}_2) = 
\frac{e^2}{\mid \br_1-\br_2 \mid}.
\end{equation}
As is well known, a contact interaction 
has to 
be handled in a truncated space defined with a cutoff energy 
$E_{\rm cut}$.
We take $E_{\rm cut}=60$ MeV, so that the s.p. space is truncated as 
\cite{EBH97}
\be
 \epsilon_{nlj}+\epsilon_{n'lj}\leq
 \frac{A_{\rm c}+1}{A_{\rm c}}\,E_{\rm cut},
\label{eq:Ecut}
\ee
where $\epsilon_{nlj}$ is a s.p. energy. 
For the Coulomb interaction, we also use the same cutoff energy. 
We have confirmed that our results do not change significantly 
even if we use a larger value of $E_{\rm cut}$ for the Coulomb interaction. 
For a given cutoff, 
the strength of the pairing interaction, $v_0$, is 
determined so as to reproduce an empirical scattering length $a$\cite{EBH97}, 
that is, 
\be
 v_0 = \frac{2\pi^2\hbar^2}{m}\frac{2a}{\pi-2ak_{\rm cut}},
\label{eq:v0}
\ee
where $k_{\rm cut}=\sqrt{{mE_{\rm cut}}/{\hbar^2}}$. 
The empirical scattering length for a neutron-neutron scattering 
is $a_{nn}=-18.5$ fm \cite{ann}, while that for a proton-proton 
scattering is $a_{pp}=-7.8063$ fm \cite{app}. 
The difference between 
$a_{nn}$ and $a_{pp}$ is mainly due to the Coulomb repulsion 
in a two-proton system. 
Since we explicitly include the Coulomb interaction in our calculations, 
we use the neutron-neutron scattering length $a_{nn}$ to determine the 
strength of the pairing interaction, Eq. (\ref{eq:v0}), assuming 
charge independence of nuclear force. 
Once $v_0$ is determined in this way, the parameters for the 
density-dependent term, $v_\rho, R_\rho$, and $a_\rho$ are adjusted 
so as to reproduce the ground state energy of $^{17}$Ne, $E_{\rm g.s.}=-0.944$ 
MeV, measured from the threshold of two-proton emission. 
See the first row in Table I for the values of the parameters. 


\begin{table}[b]
\caption{Parameter sets for the pairing interaction 
defined by Eq.(\ref{eq:Vpp}) used for the ground state of $^{17}$Ne nucleus.
$a$ is the scattering length which determines the strength of the pairing 
interaction according to Eq. (\ref{eq:v0}). 
The cutoff energy is set to be $E_{\rm cut}$=60 MeV. 
Each parameter set yields the two-proton separation energy of 
$S_{\rm 2p}=0.944$ MeV. }
\begin{center}
\begin{tabular}{c|cccc}
 \hline \hline
 & $v_0$ (MeV)
 & $v_{\rho}$ & $R_{\rho}$ (fm) & $a_{\rho}$ (fm) \\ \hline
 $a=-$18.5 fm& $-$617.46 & $-v_0$ & 3.0912 & 0.67 \\
 $a=-$15.0 fm& $-$608.37 & $-v_0$ & 3.0653 & 0.67 \\
$a=-7.81$ fm, $V^{(C)}_{\rm pp}=0$ & $-$567.73 & $-v_0$ & 3.1495 & 0.67 
\\ \hline \hline
\end{tabular}
\end{center}
\label{tb:tb2}
\end{table}

\section{results}

We now numerically solve the three-body Hamiltonian and discuss 
the ground state properties of the $^{17}$Ne nucleus. 
The results of our calculation are summarized in Table \ref{tb:tb3}.
As we mentioned in the previous section, these results were obtained 
with $a=a_{\rm nn}=-18.5$ fm. 
On the other hand, Refs. \cite{EBH97,HS05} used a somewhat different value, 
$a=-15$ fm, in order to take into account finite momentum effects. 
We have checked that the results are almost the same even if we use 
$a=-15$ fm, 
together with the parameter set for the pairing interaction shown in Table 
\ref{tb:tb2}. 

In Table 
\ref{tb:tb3},
$\langle V^{(N)}_{\rm pp} \rangle$ and $\langle V^{(C)}_{\rm pp} \rangle$ 
are expectation values of the nuclear and the Coulomb interactions, 
respectively. 
We find that the ratio of these quantities is 
about $-$0.14, that is, the Coulomb repulsion reduces 
the pairing energy by about 14\%. 
This value is similar to what has been found \cite{TD10} using a non-empirical 
pairing energy density functional for proton pairing gaps \cite{LDBM09}. 
$P([s_{1/2}]^2)$, $P([d_{5/2}]^2)$, 
and $P([d_{3/2}]^2)$ are 
the probabilities of the 
$s_{1/2}, d_{5/2}$, and $d_{3/2}$ configurations 
in the ground state wave function, respectively. 
One sees that the $d_{5/2}$ configuration dominates in the ground 
state wave function. 
$\sqrt{\langle r_{\rm NN}^2 
\rangle} \equiv 
\sqrt{\langle \Psi_{\rm g.s.} | ({\bf r}_{1}-{\bf r}_{2})^2
| \Psi_{\rm g.s.} \rangle}$ and 
$\sqrt{\langle r_{\rm C-2N}^2 \rangle} \equiv
\sqrt{\langle \Psi_{\rm g.s.} | ({\bf r}_{1}+{\bf r}_{2})^2/4
| \Psi_{\rm g.s.} \rangle} $ 
are the root-mean-square (rms) distances between the 
valence protons and that between the center of mass of the protons 
and the core nucleus, respectively. 
With these rms distances, 
the increment of the radius of $^{17}$Ne 
due to the valence protons is calculated 
as \cite{EBH97,VMP96},
\bea
\delta \left< r^2 \right> &=& \left< r^2 \right>_{A_{\rm c}+2}
    -\frac{A_{\rm c}}{A_{\rm c}+2} \left< r^2 \right>_{A_{\rm c}},
\label{eq:delr2}
\\
 &=& \frac{2A_{\rm c}}{(A_{\rm c}+2)^2} \left< r_{\rm C-2N}^2 \right>
    +\frac{1}{2(A_{\rm c}+2)} \left< r_{\rm NN}^2 \right>.
\label{eq:delr} \nonumber \\
\eea
Once 
$\delta \left< r^2 \right>$ is evaluated in this way, the matter radius 
for $^{17}$Ne can be estimated with Eq. (\ref{eq:delr2}) to be 2.62 fm. 
To this end, we used an empirical 
rms matter radius of 
the core nucleus, $r_{\rm rms}(^{15}{\rm O})$ =2.44 fm \cite{Oz01}. 
The $s$-wave probability, $P([s_{1/2}]^2)$, in our calculation 
is somewhat smaller than that in other calculations, {\it e.g.}, 
Ref. \cite{GMZ03}, and 
the rms radius for $^{17}$Ne is slightly underestimated 
as compared to the experimental value, 2.75 fm  \cite{Oz01}. 

$\left< \theta_{12} \right> $ in Table II is the opening angle between 
the valence protons. 
The value of 
$\left< \theta_{12} \right> = 76.64$ degrees is in a good agreement with 
that in Ref. \cite{BH07}. 
In Ref. \cite{BH07}, the opening angle was estimated 
using the total B(E1) value. We therefore compute also the sum rule value 
for the E1 transition (see Table I), 
\begin{equation}
 B_{\rm SR}(E1)
 = \frac{3}{\pi } Z^2_{\rm eff} e^2 \left< r_{\rm 2N-C}^2 \right>\ ,
\end{equation}
where
\begin{equation}
 Z_{\rm eff}
 = \frac{A_{\rm c}-Z_{\rm c}}{A_{\rm c}+2},
\end{equation}
for a core+2p system, $Z_{\rm c}$ 
being the proton number of the core nucleus (notice that 
$Z_{\rm eff}=Z_{\rm c}/(A_{\rm c}+2)$ for a core+2n system). 
This factor is $Z_{\rm eff}$ = 0.41 for $^{17}$Ne. 
We remind the readers that the actual value for the total B(E1) 
strength is somewhat smaller than the sum rule value due to 
the Pauli forbidden transitions \cite{EHMS07}.

\begin{table}[b]
\caption{
Ground state properties for 
${}^{17}$Ne and $^{16}$C obtained with a three-body model. 
See the text for the definition for each quantity. 
}
\begin{center}
\begin{tabular}{c|cc|c} 
\hline \hline
 & $^{17}$Ne & $^{17}$Ne (no Coul.) & $^{16}$C \\ \hline
 $\left< V_{\rm pp}^{(N)} \right>$ (MeV) & $-$3.26 & $-$2.76
 & $-$3.88  \\
 $\left< V^{(C)}_{\rm pp} \right>$ (MeV) $\quad $ & 0.448
 & 0 & 0 \\
 $P([s_{1/2}]^2)$ (\%) & 15.16 & 15.91 & 20.69 \\
 $P([d_{5/2}]^2)$ (\%) & 75.19 & 75.68 & 64.97  \\
 $P([d_{3/2}]^2)$ (\%) &  3.83 &  3.26 & 6.63 \\
 $\left< r_{NN}^2 \right> ^{1/2}$ (fm) & 4.688 & 4.749
 & 4.579 \\
 $\left< r_{C-2N}^2 \right> ^{1/2}$ (fm) & 3.037 & 3.037
 & 3.099 \\
 $\delta \left< r^2 \right> ^{1/2}$ (fm) & 1.267 & 1.273 & 1.306
 \\
 $\left< \theta_{12} \right> $ (deg) & 76.64 & 76.03 & 74.35 \\
 $B(E1)_{\rm SR}\ (e^2 {\rm fm}^2)$ & 1.494 & 1.493 & 1.290  \\
 \hline
 $r(^AZ)$ (fm) & 2.62 & 2.62 & 2.52 \\
\hline \hline
\end{tabular}
\end{center}
\label{tb:tb3}
\end{table}

It is instructive to compare our results with those obtained 
by mocking up the Coulomb interaction effectively in the density dependent 
contact interaction, {\it i.e.,} by readjusting the 
parameters for the contact interaction 
without treating the Coulomb interaction explicitly. 
For this purpose, we use the proton-proton scattering length, 
$a_{\rm pp}=-7.81$ fm, to determine the strength of the pairing interaction 
$v_0$. 
The other parameters are adjusted to reproduce the ground state energy. 
See the last row in Table \ref{tb:tb2} for the values of the parameters. 
The results are summarized in the second column in Table \ref{tb:tb3}. 
One sees that the results are almost the same between the first 
and the second columns in Table \ref{tb:tb3}. 
This indicates that the effect of Coulomb interaction on the pairing 
properties can be well simulated by adjusting the parameters for the 
pairing interaction. 

We also compare our results for $^{17}$Ne with those for $^{16}$C. 
Both nuclei have two valence nucleons outside the $N$ or $Z=8$ core, 
and one may expect some similarities between the two nuclei. 
The three-body model calculations for  $^{16}$C have been already 
performed in Refs. \cite{HS07,SMA04,HY06}. 
We here repeat the same calculation as in Ref. \cite{HS07}, but using the 
density-dependent contact interaction. 
The results for $^{16}$C are summarized in the last column in 
Table \ref{tb:tb3}. 
In order to calculate the matter radius, we use 
$r_{\rm rms}=2.30$ fm \cite{Oz01} 
for the radius of the core nucleus $^{14}$C. 
Interestingly, the ground state properties are almost the same 
between $^{17}$Ne and $^{16}$C, despite 
that the ground state energy 
for $^{16}$C ($E_{\rm g.s.}=-5.47$ MeV) 
is significantly different from that for $^{17}$Ne 
($E_{\rm g.s.}=-0.944$ MeV). 
 Both are $d$-wave dominant, and the opening angle for the valence 
nucleons is about 75 degrees. 
This similarity may be naturally understood if the 
the Coulomb interaction between 
the valence protons and the core nucleus in $^{17}$Ne 
mainly perturbs the ground state energy without affecting significantly 
the ground state wave function. 


Let us next discuss the density distribution of the valence 
nucleons, 
\be
\label{eq:rho} \rho(r_1,r_2,\theta_{12})
    =\mid \Psi_{\rm g.s.}(r_1,r_2,\theta_{12}) \mid ^2, 
\ee
which are normalized as 
\bea
&& \int_0^{\infty}4\pi r_1^2dr_1 \int_0^{\infty}r_2^2dr_2^2
     \int_0^{\pi} 2\pi \sin \theta_{12}d\theta_{12} 
\nonumber \\
 & & ~~~~~~~~~~~~~~~~~~~~~~~~~~~~~~~~\times \rho(r_1,r_2,\theta_{12}) = 1.
\eea
See Refs. \cite{BE91,HS05} for 
its explicit form. 
Figure 2 shows the density distribution for 
$^{17}$Ne (the upper panel) and 
$^{16}$C (the lower panel). 
For a presentation purpose, 
we set the radius of the valence nucleons to be the same ({\it i.e.,} 
$r_1=r_2=r$)
and multiply a weight factor of $4\pi r^2\cdot 2\pi r^2\sin\theta_{12}$.
One clearly sees that the density distributions are similar to each other 
for both nuclei. 
That is, the density distribution is largely concentrated at a small 
opening angle
$\theta_{12}\sim 20$ (deg) (notice that the density looks zero in the figure 
at $\theta_{\rm 12}=0$ because of the weight factor). 
This implies that 
the Coulomb repulsion between the valence protons plays a minor role 
in the density distribution, causing a strong diproton correlation 
in the ground state of $^{17}$Ne, as has been conjectured in 
Ref. \cite{KEHSS09} using a quasi-2D system. 

The three-peaked structure in the density distribution is due to 
the $[d_{5/2}]^2$ configuration, that 
dominates in the ground state wave function. 
Figure 3 shows the density distribution for the pure 
$[d_{5/2}]^2$ configuration for $^{17}$Ne. 
This density distribution has symmetric three peaks, which is 
given by $\rho(\theta)\propto (5/4\cdot\cos^4\theta-1/2\cdot\cos^2\theta
+3/20)$ \cite{HS07} (the formula in Ref. \cite{HS07} contains a small 
error, which is corrected here). 
With the pairing interaction, several single-particle levels 
with opposite parity are mixed up in the wave function, 
leading to the asymmetric distribution shown in Fig. 2. 

\begin{figure}[t]
\begin{center}
\includegraphics[width=7.0cm, height=5.0cm]{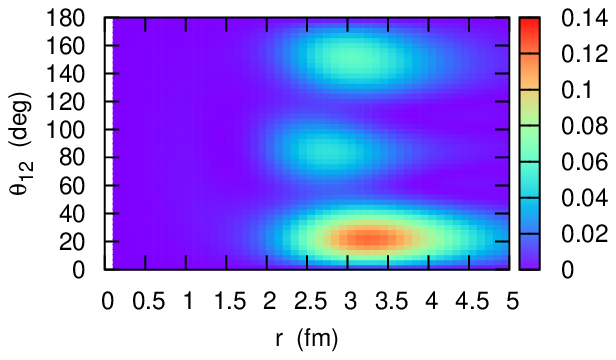}
\vspace{-0.5cm}
\includegraphics[width=7.0cm, height=5.0cm]{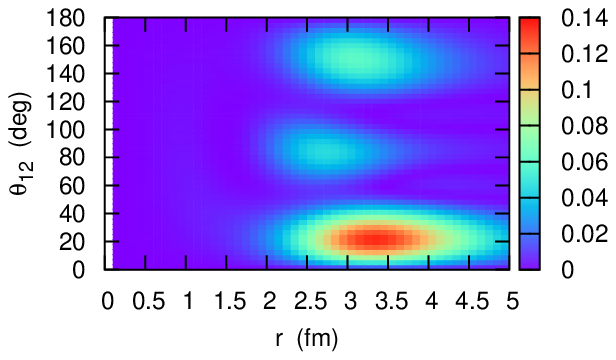}
\caption{(Color online) 
The density distribution for the ground state of $^{17}$Ne (the 
upper panel) and of $^{16}$C (the lower panel). 
It is plotted as a function of $r=r_1=r_2$ 
and the opening angle between the valence nucleons, $\theta_{12}$. 
A weight factor of $4\pi r^2\cdot 2\pi r^2\sin\theta_{12}$ has been 
multiplied. }
\end{center}
\label{fig:dens1}
\end{figure}
\begin{figure}[t]
\begin{center}
\includegraphics[width=7.0cm, height=5.0cm]{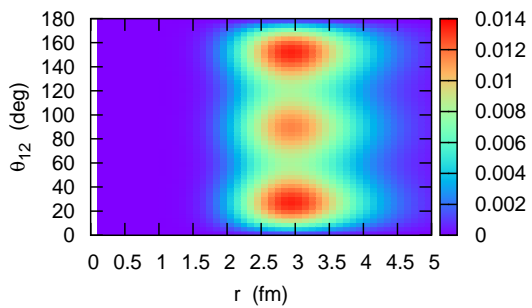}
\vspace{-0.5cm}
\caption{(Color online) 
The same as the upper panel of Fig. 2, but 
for the pure $[d_{5/2}]^2$ configuration.}
\end{center}
\label{fig:densd}
\end{figure}

We demonstrate the strong diproton correlation in $^{17}$Ne 
also in another way. 
Figure 4 shows the density distribution of the second valence nucleon 
when the first nucleon is located along the $z$ axis 
at the mean separation distance between the core and the center of mass 
of the two valence nucleons, that is, 
$r_1=\sqrt{\left< r_{\rm C-2N}^2 \right>}$. 
It is plotted as a function of the coordinate of the second nucleon, 
$z_2=r_2\cos\theta_{12}$ and $x_2=r_2\sin\theta_{12}$. 
As has been shown in Fig. 2, 
the second nucleon is well localized in the vicinity of the 
first nucleon, both for $^{17}$Ne and $^{16}$C nuclei. 
Evidently, the strong diproton correlation exists in the ground 
state of $^{17}$Ne in spite of the Coulomb repulsion between the 
valence protons.

\begin{figure}[t]
\begin{center}
\includegraphics[width=6.0cm, height=4.5cm]{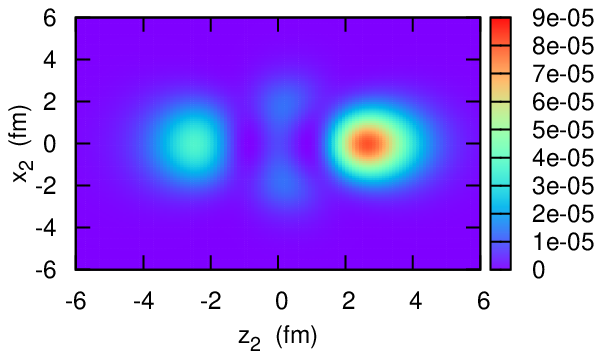}
\includegraphics[width=6.0cm, height=4.5cm]{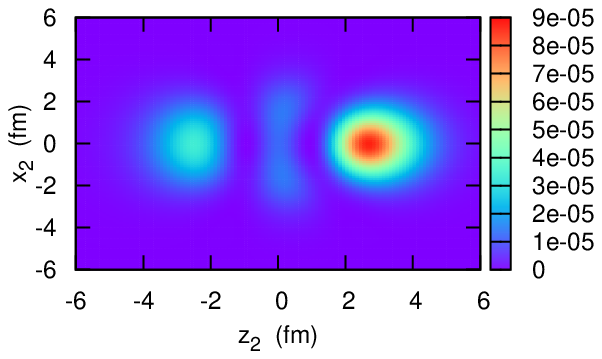}
\vspace{-0.5cm}
\caption{(Color online) The distribution of the second nucleon 
when the fist nucleon is located at 
$(z_1,x_1)=(\sqrt{\left< r_{\rm C-2N}^2 \right>},0)$. 
The upper and the lower panels are for $^{17}$Ne and $^{16}$C, respectively. }
\end{center}
\label{fig:dens2}
\end{figure}

\section{summary}

We have computed the ground state wave function for 
$^{17}$Ne based on a three-body model of $^{15}$O + $p+p$. 
To this end, we used a density-dependent contact pairing interaction 
between the valence protons. 
In addition, we explicitly treated the Coulomb interaction as well. 
We found that 
the Coulomb repulsion leads to an about 14\% reduction 
of the pairing energy. 
The density distribution for the valence protons resembles the 
two-neutron density distribution in $^{16}$C, 
showing a clear and strong 
diproton correlation in the ground state of $^{17}$Ne. 
That is, the two valence protons take a spatially compact configuration 
despite of the Coulomb repulsion. 
This suggests that the Coulomb interaction plays a minor role in 
the pairing correlation. In fact, we found that 
the effect of the Coulomb interaction can be well simulated by 
readjusting the parameters for the pairing interaction. 

The present study suggests that there is a strong diproton correlation in 
light proton-rich nuclei. It would be an interesting future work 
to study whether this 
is the case also for medium-heavy and heavy proton-rich nuclei. 
The effect of the Coulomb interaction is expected to be stronger in the 
mean-field potential for these nuclei, but it is an open question 
whether the Coulomb effect on the diproton correlation is stronger or not. 
Another future work would be to study the role of diproton correlation 
in two-proton radioactivities. 
We will report on it in a separate paper. 

\bigskip

\begin{acknowledgments}
This work was supported
by the Grant-in-Aid for Scientific Research (C), Contract No.
22540262 and 20540277 from the Japan Society for the Promotion of Science.
\end{acknowledgments}

\end{document}